\documentclass[structabstract]{aa}  
\usepackage{graphicx}
\usepackage{txfonts}
\usepackage{natbib}

\begin{document}
   \title{The Doppler Shadow of WASP-3b}

   \subtitle{A tomographic analysis of Rossiter-McLaughlin observations}

   \author{G. R. M. Miller\inst{1}
          \and
          A. Collier Cameron\inst{1}
	  \and
	  E. K. Simpson\inst{2}
          \and
          D. Pollacco\inst{2}
          \and 
          B. Enoch\inst{1}
          \and
          N.P. Gibson\inst{3}
          \and
          D. Queloz\inst{4}
          \and
          A.H.M.J. Triaud\inst{4}
          \and
          G. H\'ebrard\inst{5}
          \and
	  I. Boisse\inst{5}
	  \and
          C. Moutou\inst{6}
          \and
          I. Skillen\inst{7}
          }

   \institute{School of Physics \& Astronomy, University of St Andrews,
              North Haugh, St Andrews, Fife KY16 9SS, UK\\
              \email{gm228@st-andrews.ac.uk}
         \and
             Astrophysics Research Centre, School of Mathematics \& Physics,
             Queen's University, University Road, Belfast, BT7 1NN, UK
         \and
             Department of Physics, University of Oxford, Denys Wilkinson Building, Keble Road, Oxford OX1 3RH, UK
         \and
             Observatoire de l'Universit\'e de Gen\`eve, Chemin des Maillettes 51, CH-1290 Sauverny, Switzerland	
	 \and
             Institut d'Astrophysique de Paris, UMR7095 CNRS, Universit\'e Pierre \& Marie Curie, 98$^{bis}$ bvd. Arago, 75014 Paris, France
         \and
             Laboratoire d’Astrophysique de Marseille, BP 8, 13376 Marseille, Cedex 12, France
	 \and
             Isaac Newton Group of Telescopes, Apartado de Correos 321, E-38700 Santa Cruz de la Palma, Tenerife, Spain}

   \date{Received May 27, 2010; accepted August 18, 2010}

 
  \abstract
   {Hot-Jupiter planets must form at large separations from their host stars
   where the temperatures are cool enough for their cores to condense. They
   then migrate inwards to their current observed orbital separations. Different theories of how this migration occurs
   lead to varying distributions of orbital eccentricity and the alignment between the rotation axis of the star and the orbital axis of the planet.}
   {The spin-orbit alignment of a transiting
   system is revealed via the Rossiter-McLaughlin effect, which is the anomaly present in the radial velocity measurements of the rotating star 
   during transit due to the planet blocking some of the starlight. In this paper we aim to measure the spin-orbit alignment of the WASP-3 system via a new way of analysing the Rossiter-McLaughlin observations.}
   {We apply a new tomographic method for analysing the time variable asymmetry of stellar line profiles caused by
   the Rossiter-McLaughlin effect. This new method eliminates the systematic
   error inherent in previous methods used to analyse the effect.}
   {We find a value for the projected stellar spin rate of $v\sin i=13.9\pm 0.03$ km s$^{-1}$ which is in agreement with previous measurements but          has a much higher precision. The system is found to be well aligned, with $\lambda = 5_{-5}^{+6}$$^\circ$ which favours an evolutionary history for WASP-3b involving
    migration through tidal interactions with a protoplanetary disc. From comparison with isochrones we put an upper limit on the age of the star of 2 Gyr. }
  {}

\maketitle

%

\section{Introduction}

   Since the confirmation in 1995 of the first extrasolar planet orbiting
   a main sequence star \citep{mayor1995} it has become evident that gas-giant planets
   with orbital separations smaller than that of Mercury are quite common. 
   In fact, these Hot Jupiters account for around 20\% of the approximately 450\footnote{For up-to-date exoplanet statistics see http://www.exoplanet.eu} 
   exoplanets discovered to date, though their current prevalence in transit searches is mostly due to observational bias and recent results from the Kepler Mission suggest that the majority of exoplanets are of lower mass \citep{borucki2010}.
   It has been shown that these
   planets must form farther out, beyond what is known as the `snow line' \citep{sasselov2000},
   and then migrate inwards to their current locations \citep{lin1996}. If this is the case 
   then through what process does this migration occur and how can we gain
   information on this from current observations?

   There are three main competing theories of Hot-Jupiter migration. The first
   states that migration occurs due to tidal interactions between the planet and the protoplanetary disc
   causing some of the planet's angular momentum to be lost to the disc \citep{goldreich1980, nelson2000}. The second
   proposes that gravitational scattering between multiple planets in a system
   can cause one planet to migrate inwards at the expense of the other being shot out of the system \citep{weidenschilling1996, chatterjee2008, juric2008}.

   These separate scenarios should lead to quite different post-migration states. Tidal
   interactions between the planet and protoplanetary disc will result in a system where the orbital
   axis of the planet is well aligned with the rotation axis of the host star.
   It would also lead to highly circularised orbits. The planet-planet scattering process
   on the other hand would result in planets which have larger orbital eccentricities
   and are not necessarily well aligned \citep{weidenschilling1996}.

   The third proposed process involved in Hot Jupiter migration is
   the Kozai mechanism. It involves oscillations between the spin-orbit alignment angle
   and the orbital eccentricity of the planet due to the presence of another (outer)
   planet in the system or a binary companion of the host star \citep{kozai1962, wu2003, fabrycky2007}. During these
   oscillations the value $\sqrt{1-e^2}\cos i$ is conserved, where $e$ is the planet's
   orbital eccentricity and $i$ is its inclination to the orbital plane of the other
   objects. \cite{nagasawa2008} suggested a combination of all three processes
   could be responsible for the migration of Hot Jupiters.

   Measuring the alignments of the known Hot-Jupiter systems
   helps shed some light on the migration process and refine
   models of system evolution. The misalignment angle $\lambda$ between
   the planet's orbital axis and the rotation axis of the host star can
   be determined via the Rossiter-McLaughlin effect \citep{rossiter1924, mclaughlin1924}. This is the radial velocity 
   anomaly observed during transit due to the planet blocking some of the starlight.
   A planet in a prograde orbit would first block some blue-shifted light,
   from the half of the star which is rotating towards us, causing an anomalous
   red-shift in the star's radial velocity. Then once it has crossed the stellar
   rotation axis it will block light from the receding half of the disc causing
   an anomalous blue-shift. The form of this radial velocity anomaly allows us to measure
   the projected stellar rotation rate and the spin-orbit misalignment angle \citep{gaudi2007}. 

   \cite{ohta2005} and \cite{gimenez2006} derived detailed expressions for measuring the anomalous shift
   in the line centroid during transit. Using the radial velocity shift
   to determine the parameters related to the Rossiter-McLaughlin effect can be a
   problem when the spectrograph can resolve the stellar line profile. The data pipelines for instruments
   such as SOPHIE and HARPS calculate RVs by fitting a Gaussian profile to the cross-correlation function (CCF) in order
   to measure the shift in line centroid. However, as the planet is blocking some of the
   light from the star this shows up on the line profile as a travelling 'bump' of width
   equal to that of the local non-rotating line profile. This means that the CCF is
   now asymmetric and time-varying during transit, introducing a systematic deviation to measurements
   of the radial velocity \citep{winn2005}. This error becomes greater for more rapidly rotating host stars
   as their line profiles exhibit higher degrees of rotational broadening \citep{hirano2009}.

   In more recent studies \cite{winn2005, winn2006, winn2007} tried to account for this by building models of the
   out of transit line profile and the light blocked by the planet. They included these into
   their analysis of the line-spread function and produced semi-empirical corrections to the
   model radial velocities. However, their method does not account for the problem entirely as their analysis of HD 189733b shows a clear pattern
   of correlated residuals in the radial velocities during transit \citep{winn2006}. A similar pattern in the
   radial velocity residuals of HD 189733b was found by \cite{triaud2009} who discuss their cause
   in detail. \cite{simpson2010} also found a similar correlation pattern when analysing the WASP-3 system
   using the expressions from \cite{ohta2005} with the corrections developed by \cite{hirano2009}. 
   In order to eliminate the need for empirical corrections \cite{cameron2009} developed a new method that involves decomposing
   the CCF profile into its various components, namely the limb-darkened rotation profile, gaussian average line profile and the travelling signature caused by the transiting planet.

   To gain information on the distribution of spin-orbit misalignment angles, the WASP (Wide Angle Search for Planets) consortium
   has been performing follow-up spectrographic observations of the transits of the
   known WASP planets. Observations have been made using the HARPS spectrograph on the ESO 3.6m telescope at La Silla and
   the SOPHIE spectrograph on the 1.93m telescope at the
   Observatoire de Haute-Provence. WASP-3b is the third planet discovered through the SuperWASP project \citep{wasp3a}.
   It is a Hot Jupiter orbiting a main sequence star of spectral type F7-8V.

   In this study we implement the method of \cite{cameron2009} to remove the inherent error present in the Gaussian-fitting
   method by instead focusing on the light removed by the planet. We model all the components
   of the line-spread function and fit to spectral observations of WASP-3b in order to track the trajectory
   of the missing light component as it crosses the line profile during transit. A model of
   the out-of-transit profile is subtracted from the data leaving only the signature of the 
   blocked light. The trajectory of this feature is used to derive values for 
   the projected stellar rotation rate, the impact parameter and the spin-orbit misalignment.

   This method does not rely on high precision radial velocity measurements, making it uniquely useful for detecting planets orbiting rapid-rotating early-type stars \citep{wasp33}. Many of these stars which have reliable photometric data have been dropped from radial velocity follow-up observations due to their rotation rate and line-poor spectra.

\section{Observations and analysis}

   We reanalyse data presented in \cite{simpson2010} who used the SOPHIE echelle spectrograph \citep{bouchy2009} on the 1.93m telescope at the
   Observatoire de Haute-Provence to take 26 observations
   before, during and after the transit of WASP-3b on the night of
   September 30th 2008. A major advantage of this new method is that it does not require high precision radial velocity measurements, therefore the spectrograph was used in high efficiency mode ($R=40000$). The exposure times for the observations ranged from 300 - 1800 seconds to ensure a constant signal-to-noise ration of 35. In total 137 minutes of observations were taken during transit and 130 minutes out of transit.
   CCFs were computed using
   the automated SOPHIE data-reduction pipeline which is adapted
   from the HARPS data-reduction software. WASP-3 is of spectral type
   F7-8V, so the weighted mask function used for the cross-correlation
   was for a G2V star as this was the closest type offered by the pipeline.
   For a more detailed explanation of the observations and data reduction procedure see \cite{simpson2010}.

\subsection{Modelling the observations}

   The following sections outline the key points of the new method used to analyse the CCF data.
   The modelling process is described in detail by \cite{cameron2009}.
   
   First we need to build a model of the out-of-transit CCF. This is done by
   convolving a Gaussian representing the the local
   line profile at any point on the surface of the star with a limb-darkened rotation profile.
   For the local line profile we use

   \begin{equation}
   g(x)=\frac{1}{\sqrt{2\pi}s}e^{-\frac{x^2}{2s^2}}
   \end{equation}
   For the limb-darkened rotation profile we use
   the equation

   \begin{equation}
   f(x)=\frac{6((1-u)\sqrt{1-x^2}-\pi u(x^2-1)/4)}{\pi (u-3)}
   \end{equation}

   assuming a linear limb-darkening model where $u$ is the limb-darkening coefficient and $-1<x<1$.
   We adopted a value of $u=0.69$ taken from the \cite{claret2004} tables for the g' filter.
   This value corresponds to a star with $T_{eff}=6500$ K, $\log g_*=4.5$ and $[M/H]=0$ which best describe
   the values for WASP-3 calculated through analysis of the SOPHIE spectra and presented in the discovery paper ($T_{eff}=6400\pm{100}$K, $\log g_*=4.5\pm{0.05}$). 

   The convolution of these two functions is given by

   \begin{equation}
   h(x)=\int\limits_{-1}^1 f(z)g(x-z)dz
   \end{equation}

   which is calculated by numerical integration.
   We need to shift the model CCF to account for the fact
   that the CCFs are computed in the velocity frame of the
   solar system barycentre. To do this we compute

   \begin{equation}
   x_{ij}=v_{ij}-(K(e\cos\omega + \cos(\nu_j +\omega))+\gamma)
   \end{equation}

   where $v_{ij}$ is the velocity of pixel $i$ in the barycentric frame, $\nu_j$ is the true anomaly at the time of the $j$th observation, $\omega$ is the argument of periastron, $e$ is the orbital eccentricity, $K$ is the radial velocity amplitude and $\gamma$ is the systemic centre-of-mass velocity.

   At any moment in time the position of the planet on the plane of the sky
   is given by the co-ordinates

   \begin{equation}
   x_p=r\sin(\nu + \omega - \pi/2)
   \end{equation}
   \begin{equation}
   z_p=r\cos(\nu + \omega - \pi/2)\cos i
   \end{equation}

   where $r$ is the instantaneous distance of the planet from the star
   and $i$ is the inclination of the orbital axis to
   the line-of-sight.
   The perpendicular distance of the planet from the stellar rotation axis
   in units of $R_*$ is now

   \begin{equation}
   u_p=x_p\cos\lambda-z_p\sin\lambda
   \end{equation}

   where $\lambda=\phi_{spin}-\phi_{orbit}$ and $\phi$ is the position angle in the
   plane of the sky \citep{winn2005}.
   Combining the model out-of-transit CCF with the model of the missing starlight
   gives us

   \begin{equation}
   p_{ij}=h(x_{ij})+\beta g(x_{ij}-u_p)
   \end{equation}
   
   Here the term $h(x_{ij})$ is the model of the out-of-transit stellar CCF.
   The term $\beta g(x_{ij}-u_p)$ represents the travelling Gaussian signature
   caused by the missing starlight, where $\beta$ is the fraction of starlight blocked
   by the planet during the total part of the eclipse.
   Fig.2 shows the original CCFs and the resulting residuals when first the
   model CCFs $h(x_{ij})$ are subtracted off leaving the signature of the missing
   starlight and secondly the overall residuals when the complete model $p_{ij}$ is subtracted.

\subsection{Fitting the model}

   In order to orthogonalise the data and the model we subtract their optimal
   mean values using inverse-variance weights $w_{ij}=1/\sigma^2_{ij}$:

   The goodness of fit of the model to the observed data is then calculated
   using the $\chi^2$ statistic

   \begin{equation}
   \chi^2=\sum_{i=1}^{n}(d'_{ij}-\hat{A}p'_{ij}-\alpha_i)^2\omega_{ij}
   \end{equation}   
 
   where $\alpha_i$ is the optimal average of the residual spectra and $\hat{A}$ is
   a multiplicative constant calculated via optimal scaling.

   The spectral resolving power of the SOPHIE spectrograph is R=75000 which
   gives a velocity resolution of 4 km s$^{-1}$. However, the CCFs are produced
   with velocity increments of 0.5 km s$^{-1}$ meaning that the errors on the data
   from neighbouring pixels are correlated. In order to make sure our
   data points were statistically independent we binned them by a factor of 8 before
   computing $\chi^2$.

\subsection{Markov chain Monte Carlo technique}

The model parameters were calculated using
a Markov chain Monte Carlo (MCMC) method. The MCMC code used for this study is a hybrid 
of the code previously used to calculate the parameters of the WASP systems from photometric
and spectroscopic data sets \citep{cameron2007}, and a new code developed by \cite{cameron2009} specifically for RM analysis.
Therefore at the same time as calculating the parameters from the RM effect we recalculated all the MCMC fitting parameters for the system.
A comparison of our results with those of the initial MCMC analysis from \cite{wasp3a} can be seen in Table 1.
The code includes the mass and radius calibration \citep{torres2010} which was recently implemented in the MCMC analysis by \cite{enoch2010}. The method described by \cite{torres2010} is used to derive the stellar mass and radius from polynomial functions
of $T_{eff}$, $\log g_*$ and [Fe/H].
As in \cite{cameron2009} we replaced the width of the local non-rotating profile $s$ with $v_{CCF}$, the FWHM of
the CCF in an attempt to avoid correlated pairs of parameters. 

\begin{equation}
v_{CCF}=\sqrt{(v\sin i)^2+v_g^2}
\end{equation}

where $v_g$ is the FWHM of the Gaussian representing the local stellar and instrumental line profile.

\begin{equation}
v_g=2sv\sin i\sqrt{\ln 2}
\end{equation}

At each step in the chain the current values of the four parameters are altered by a Gaussian perturbation
and then the goodness-of-fit is recalculated. For example, the proposed next step for $\lambda$ 
is

\begin{equation}
\lambda_k=\lambda_{k-1}+f\sigma_{\lambda}G(0,1)
\end{equation}

where $f$ is a scale factor of order unity and $G(0,1)$ is a random number drawn from a Gaussian distribution of zero mean and unit variance.
Steps are accepted or rejected in accordance with 
the Metropolis-Hastings algorithm. After each proposed step $\Delta\chi^2=\chi_k^2-\chi_{k-1}^2$ is calculated.
If $\Delta\chi^2<0$ then the proposed step is accepted. If $\Delta\chi^2>0$ then the step is accepted
with a probability of $e^{-{\Delta\chi^2/2}}$. Each successful step is recorded to the MCMC output. If
a step is rejected then the previous accepted step is recorded again. Varying the scale factor $f$ changes
the acceptance rate of the MCMC.
We ran the MCMC with an initial burn-in phase of 1000 steps. After the first burn-in period we re-evaluated our
estimates of the variances on the binned CCF data. We then ran the chain for another 100 steps in order
to re-evaluate the variances on the four fitting parameters from the chains. This was followed by a final production run of 10,000 steps
where $f$ was fixed at a value of 0.5 as this was found to return the desired acceptance rate of 25\%.

   \begin{figure*}[htp]
   \begin{center}
   \includegraphics[width=15cm]{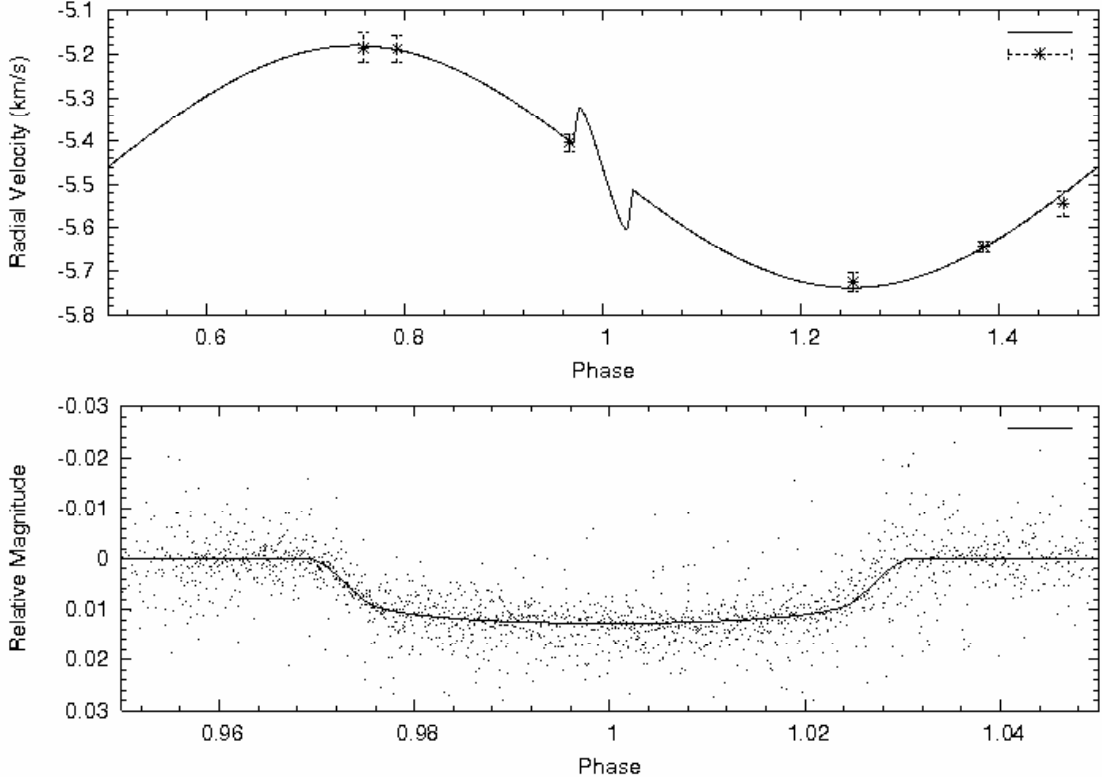}
      \caption{Upper panel: Phase-folded plot of the 6 out-of-transit radial velocity measurements which are not affected by any time-varying asymmetry of the line-profile. The out-of-transit RV fit was calculated using the velocity semi-amplitude, orbital eccentricity, arguement of periastron and true anomaly. The position of the planet over the stellar disc, ratio of star/planet radii, impact parameter and non-linear limb darkening coefficients were used to model the RM anomaly during transit. Lower panel: Phase-folded plot of all 8 sets of photometric data analysed in this study.
              }
   \end{center}
         
   \end{figure*}

\subsection{Photometric and spectroscopic datasets}

In total we used 8 photometric and 2 radial velocity datasets in the MCMC analysis. In addition to the photometric datasets analysed
in the discovery paper we included 2 additional sets of photometry taken with the RISE instrument on the 2m Liverpool Telescope
at Observatorio del Roque de Los Muchachos, La Palma \citep{gibson2008}. The radial velocity data comprised the 26 observations described earlier 
to target the RM effect, and the 6 out-of-transit observations taken in July and August 2007 also using the SOPHIE spectrograph
as described by \cite{wasp3a}. Figure 3 shows phase-folded plots of the photometric data and the out-of-transit radial velocity measurements.

   \begin{figure}[htp]
   \centering
   \includegraphics[width=9cm]{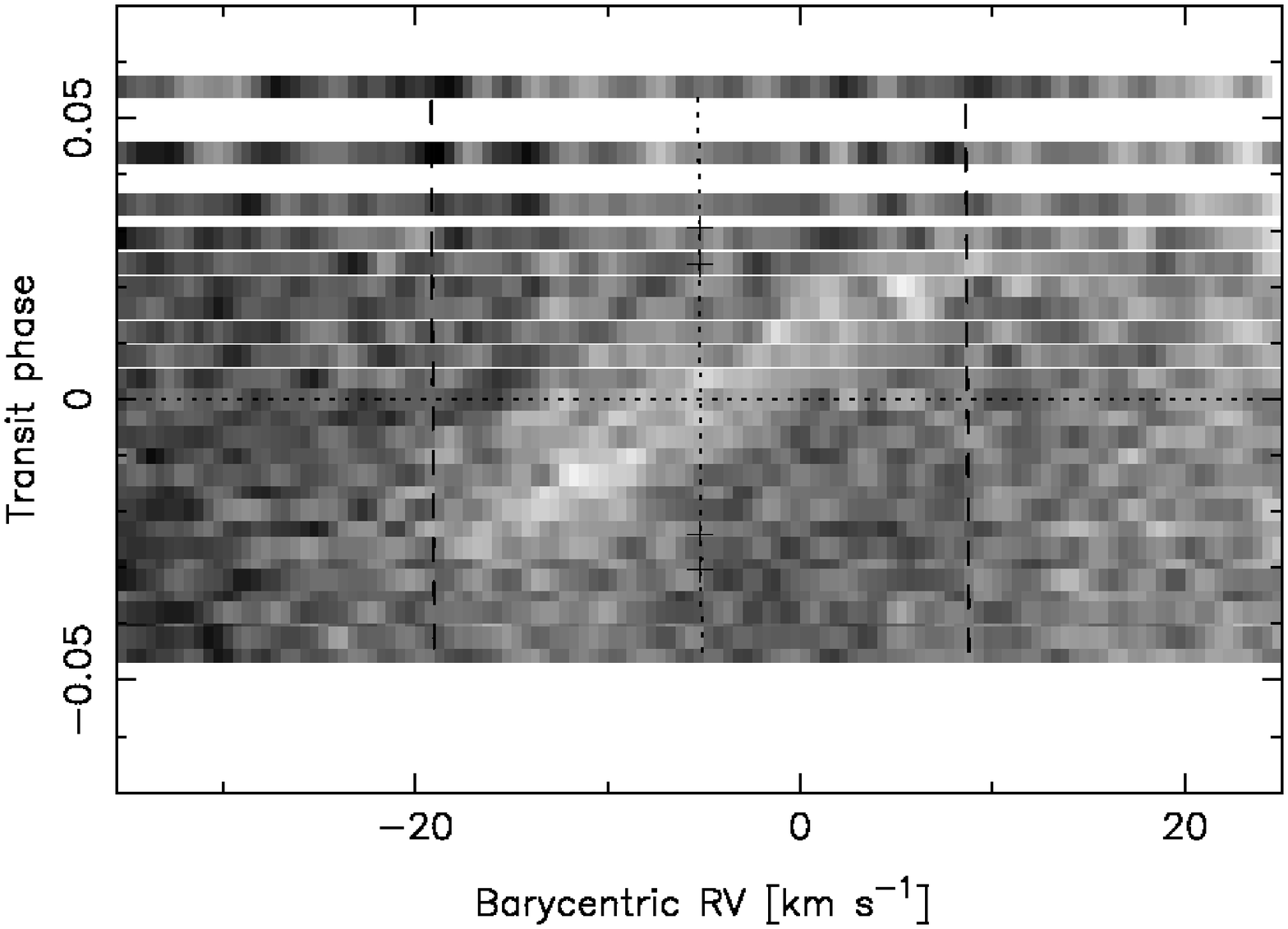}
   \includegraphics[width=9cm]{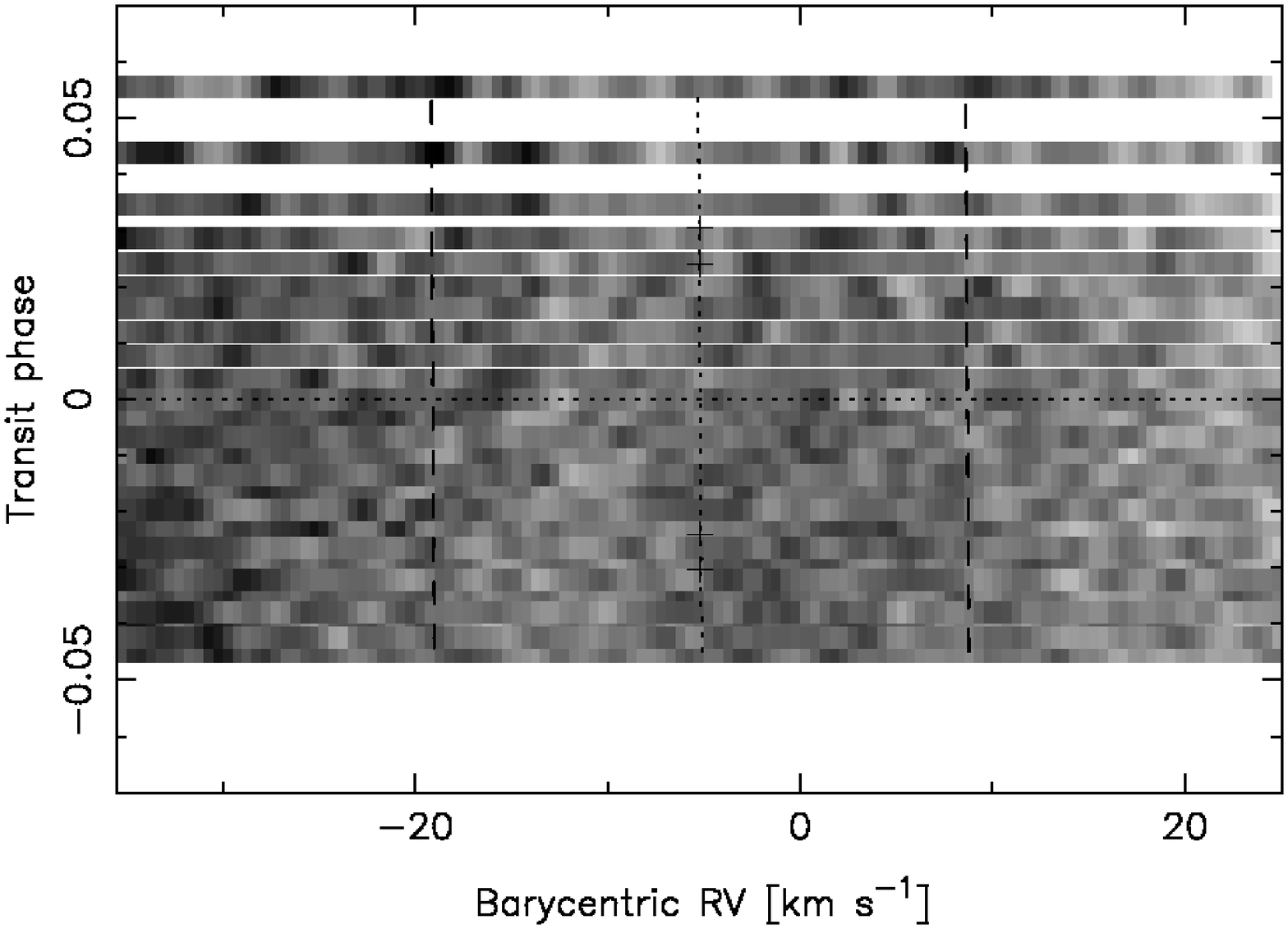}
   \caption{\emph{Top}: Residual map of time series CCFs with the model spectrum subtracted leaving the bright time-variable feature due the light blocked
            by the planet. \emph{Bottom}: Here the best-fit model for the time-variable feature
            has also been removed to show the overall residual. The horizontal dotted line marks the phase of mid-transit. The shift of this line from zero shows the underlying systemic radial velocity. The two vertical dashed lines are at $\pm{v\sin i}$ from the stellar radial velocity (marked by the vertical dotted line). The crosses on the vertical dotted line denote the two points of contact at both ingress and egress.}
              
    \end{figure}
%

   \begin{table*}[tbp]
      \caption[]{Results of MCMC analysis compared to those presented in the literatures.}
         
     $$ 
         \centering
         \begin{tabular}{lccccl}
            \hline
            \noalign{\smallskip}
            Parameter & Symbol & Pollacco et al. (2007) & Gibson et al. (2008) & This study & Units \\
            \noalign{\smallskip}
            \hline
            \noalign{\smallskip}
            Transit epoch (JD-2450000)  & $T_{0}$  &  $4143.8503_{-0.0003}^{+0.0004}$  & $4605.5592_{-0.0002}^{+0.0002}$ &  $3992.4101_{-0.0004}^{+0.0004}$       & days    \\
	    Orbital period  & P  &  $1.846834_{-0.000002}^{+0.000002}$ & $1.846835_{-0.000002}^{+0.000002}$ &  $1.846837_{-0.000001}^{+0.000001}$   & days   \\

            Planet/star radius ratio  & $(R_{P}/R_{*})^2$ &  $0.0106_{-0.0004}^{+0.0002}$  & $0.0103_{-0.0001}^{+0.0002}$ & $0.0105_{-0.0001}^{+0.0002}$  &   \\
            Transit duration          & $t_T$       &  $0.1110_{-0.0018}^{+0.0009}$      &$0.1147_{-0.0005}^{+0.0008}$&  $0.1126_{-0.0006}^{+0.0007}$ & days\\
            Impact parameter          & b         &  $0.505_{-0.166}^{+0.051}$         &$0.448_{-0.014}^{+0.014}$& $0.38_{-0.07}^{+0.11}$     & R$_{*}$  \\ [6pt]
            
            Stellar reflex velocity   & K$_{s}$     &  $0.2512_{-0.0108}^{+0.0079}$ &-&  $0.2782_{-0.0134}^{+0.0138}$   &  km s$^{-1}$   \\
            Centre-of-mass velocity   & $\gamma$  & $-5.4887_{-0.0018}^{+0.0013}$ &-&  $-5.4599_{-0.0036}^{+0.0037}$  &  km s$^{-1}$ \\
            Orbital semi-major axis   & a         &  $0.0317_{-0.0010}^{+0.0005}$ &-& $0.0313_{-0.0001}^{+0.0001}$  &  AU \\
            Orbital inclination       & i         &  $84.4_{-0.8}^{+2.1}$         &$85.06_{-0.15}^{+0.16}$&  $87.0_{-1.1}^{+1.0}$ &  degrees  \\ [6pt]

	    Stellar mass              & M$_{*}$     &  $1.24_{-0.11}^{+0.06}$   &-&  $1.20_{-0.01}^{+0.01}$     & M$_{\odot}$  \\
	    Stellar radius            & R$_{*}$     &  $1.31_{-0.12}^{+0.05}$   &-&  $1.21_{-0.03}^{+0.04}$     & R$_{\odot}$ \\
            Stellar surface gravity   & $\log g_*$&  $4.30_{-0.03}^{+0.07}$   &-&  $4.33_{-0.03}^{+0.03}$     & [cgs]   \\
            Stellar density           & $\rho_*$  &  $0.55_{-0.05}^{+0.15}$   &-&  $0.67_{-0.06}^{+0.05}$     & $\rho_{\odot}$   \\ [6pt]

            Planet mass               & M$_p$       & $1.76_{-0.14}^{+0.08}$ &$1.76_{-0.14}^{+0.08}$&   $1.90_{-0.09}^{+0.10}$  & M$_J$    \\
            Planet radius             & R$_p$       & $1.31_{-0.14}^{+0.07}$ &$1.29_{-0.12}^{+0.05}$&   $1.20_{-0.03}^{+0.05}$       & R$_J$    \\
            Planetary surface gravity & $\log g_p$& $3.37_{-0.04}^{+0.09}$ &$3.42_{-0.04}^{+0.06}$&   $3.47_{-0.04}^{+0.03}$      & [cgs]    \\
            Planet density            & $\rho_p$  & $0.78_{-0.09}^{+0.28}$ &$0.82_{-0.09}^{+0.14}$&   $1.08_{-0.13}^{+0.11}$      & $\rho_J$ \\
            Planet temp               & T$_{eql}$   & $1960_{-76}^{+33}$ &-&   $1920_{-22}^{+32}$    & K    \\ 
            \noalign{\smallskip}
            \hline
         \end{tabular}
     $$
   \end{table*}

   \begin{table*}[tbp]
      \caption[]{Values for projected stellar rotation rate, spin-orbit misalignment angle and width of the intrinsic stellar line profile compared to those from the two previous Rossiter-McLaughlin studies of WASP-3b. Note: The value obtained for $v\sin i$ through spectral analysis is $13.4\pm{1.5}$ km s$^{-1}$ \citep{wasp3a}. }
         
     $$ 
         \begin{tabular}{lcccl}
            \hline
            \noalign{\smallskip}
            Parameter & Simpson et al. (2009) & Tripathi et al. (2010) & This study & Units \\
            \noalign{\smallskip}
            \hline
            \noalign{\smallskip}
            $v\sin i$ & $15.7_{-1.3}^{+1.4}$ & $14.1_{-1.3}^{+1.5}$ & $13.9_{-0.03}^{+0.03}$ & km s$^{-1}$  \\
            $\lambda$ & $13_{-7}^{+9}$ & $3.3_{-4.4}^{+2.5}$ & $5_{-5}^{+6}$    & degrees    \\
            \noalign{\smallskip}
            \hline
         \end{tabular}
     $$ 

   \end{table*}
%

   \begin{figure*}
   \begin{center}
   \includegraphics[width=15cm]{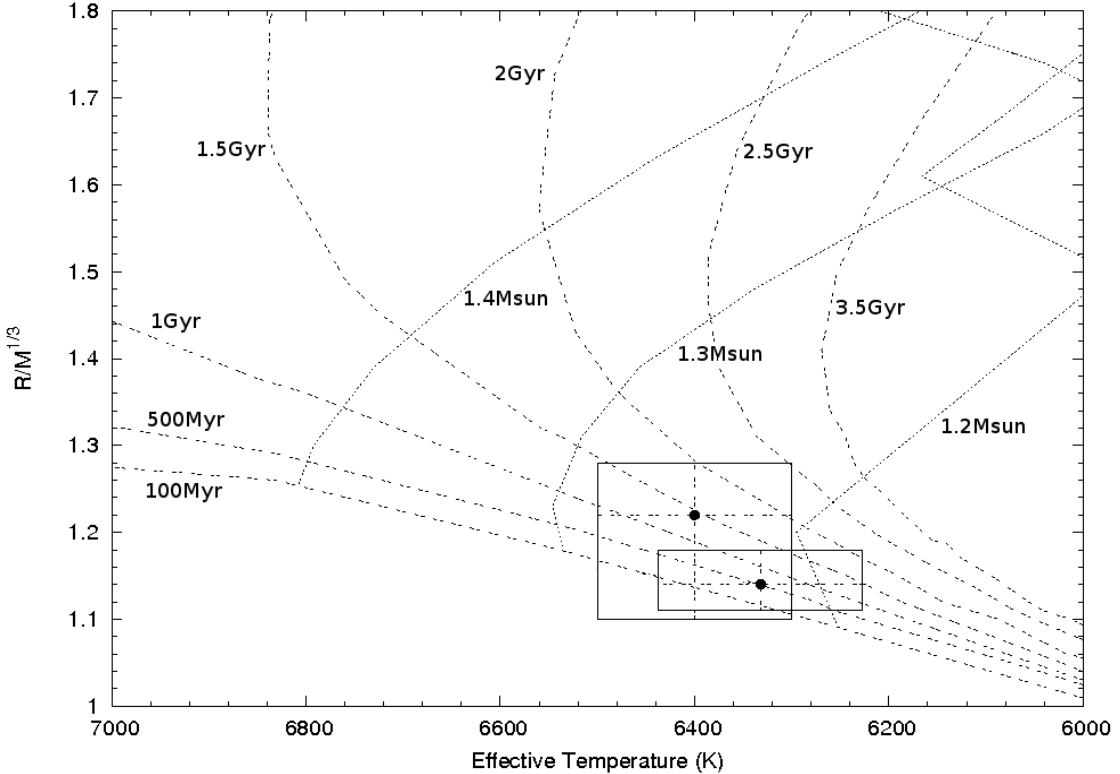}
      \caption{The new position of WASP-3 in the $T_{eff}$ vs $R/M^{1/3}$ plane. The larger of the two boxes shows the data from the discovery paper. The smaller box shows the error range from this study. The lines show evolutionary tracks from Girardi et al. (2000) for 1.2, 1.3 and 1.4 solar masses and isochrones at 0.1, 0.5, 1, 1.5, 2, 2.5 and 3 billion years. The tracks and isochrones here are for stars with solar metalicity, $[M/H]=0$. 
              }
   \end{center}	
         
   \end{figure*}
%

   \begin{figure*}[htp]
   \begin{center}
   \includegraphics[width=15cm]{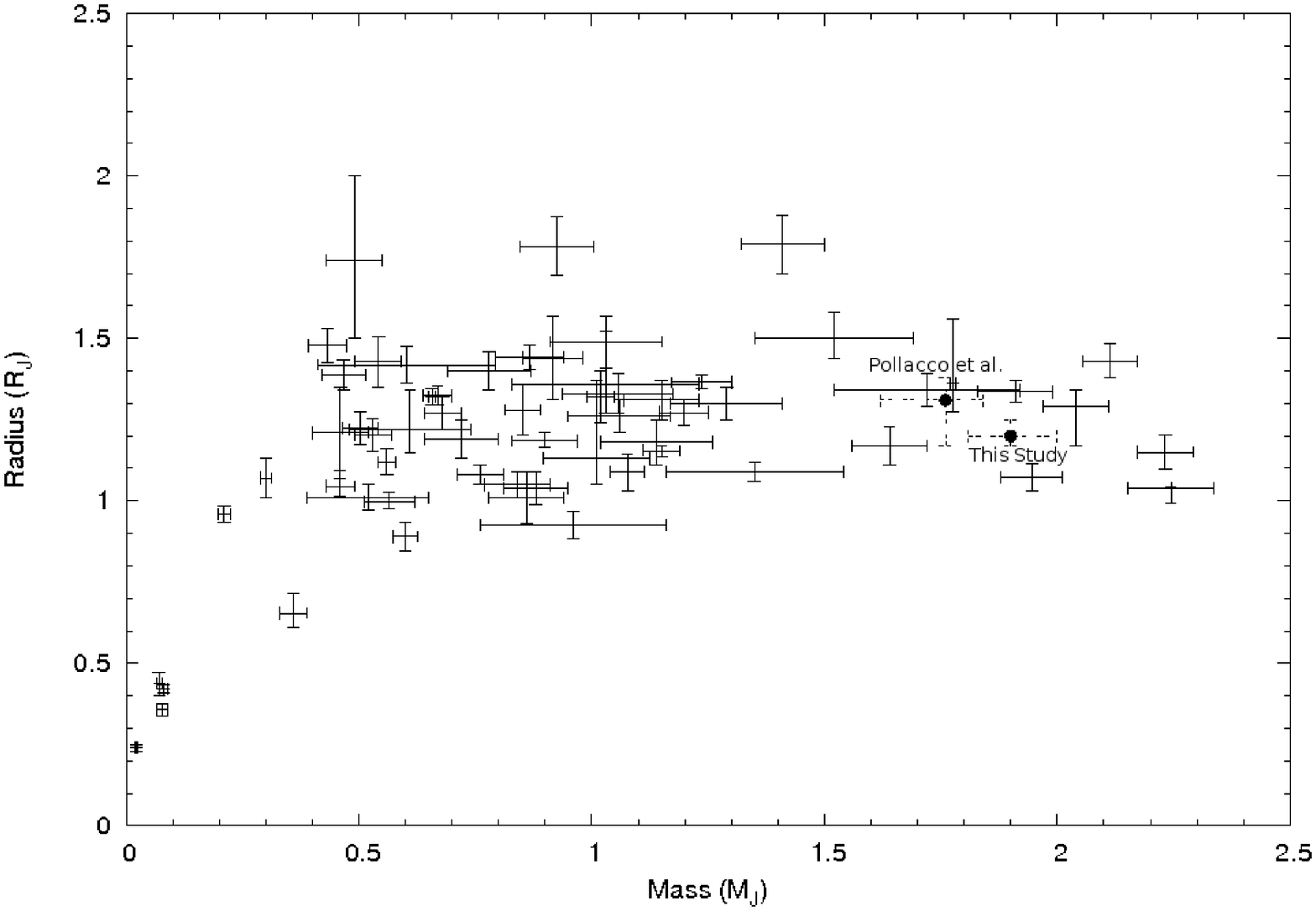}
      \caption{Mass vs Radius plot for known transiting planets. The two black dots indicate the previous position of WASP-3b from the results in the discovery paper and its updated position from the mass and radius values determined in this study.  
              }
   \end{center}	
         
   \end{figure*}

\section{Results}

   The resulting values for the system parameters can been seen in Table 1
   alongside those taken from the discovery paper, both with their $1\sigma$
   errors. With the exception of the impact parameter, all values are in agreement
   within their $1\sigma$ uncertainty ranges. Through examination of its relation with the stellar surface gravity \cite{wasp3a} suggest that the impact
   parameter should lie between 0.4 and 0.6 which is in good agreement with
   the value found in this study of $b=0.38_{-0.07}^{+0.11}$. Using our
   method the impact parameter can be more closely constrained, as the properties of the streak of 'missing light' also give us a measure of the latitudes
   of ingress and egress.

   One major success of this study is that sensible values in agreement
   with previous work are produced without having to fix any of the 
   system parameters in place. \cite{wasp3a} showed that in order
   to reconcile the MCMC analysis with spectroscopic diagnostics $\log g_*$
   must be between 4.25 and 4.35. Recently \cite{enoch2010} attempted
   to reproduce the results from the WASP-3 discovery paper incorporating the Torres mass calibration
   into the MCMC analysis and found they also needed to fix $\log g_*$ to the value determined by spectroscopic analysis.
   In this study we left $\log g_*$ as a floating variable and found its value to be $\log g _* = 4.33_{-0.03}^{+0.03}$ 
   which lies within the range suggested in the discovery paper.

   The results produced for the RM parameters
   are presented in Table 2 along with their one sigma errors.
   The value we obtained for the projected stellar rotation rate $v\sin i=13.9\pm{0.03}$ km s$^{-1}$ is in good agreement with
   the value of $v\sin i=13.4\pm{1.5}$ km s$^{-1}$ derived by the analysis
   of the SOPHIE spectroscopy as presented in the WASP-3 discovery paper \citep{wasp3a}.
   However, our result shows a much greater level of precision. This is because uncertainties on previous estimates
   of $v\sin i$ are removed by the fact that we can measure the FWHM of the intrinsic profile directly.
   Our analysis finds the projected spin-orbit misalignment angle to be $\lambda =5_{-5}^{+6}$$^\circ$, which is almost
   indistinguishable from zero.

\cite{simpson2010} recently performed a Rossiter-McLaughlin effect analysis of the WASP-3 system with the same data used in this study. They
used the method of \cite{hirano2009} to account for the aforementioned systematic error encountered
when trying to calculate $v\sin i$ by applying the \cite{ohta2005} method. Their value for $v\sin i$ of $15.7_{-1.4}^{+1.3}$ km s$^{-1}$ is similar   to our 
value but still slightly larger. For $\lambda$ they derived a value of $13_{-7}^{+9}$$^\circ$ which agrees
with our result. In addition to this \cite{tripathi2010} performed an analysis of a separate set of data and retrieved values that are in good       agreement with those found in this study, as can be seen in Table 2.

   Using our new, more accurate values for the system parameters we were able to plot the position of WASP-3 on the $R/M^{1/3}$ vs $T_{eff}$ plane (see Fig. 3). From the MCMC analysis we obtained a value for the stellar effective temperature of $T_{eff}=6332\pm{105}$ K. Comparing this temperature range with the evolutionary tracks from \cite{girardi2000} we can get a separate mass estimate for the star of $M_{*}=1.23\pm{0.04}$ M$_{\odot}$. This is in good agreement with our previous value of $M_{*}=1.24_{-0.11}^{+0.06}$ M$_{\odot}$. From the maximum error range we can put an upper limit on the stellar age of around 2 Gyr. This is an improvement on the value presented in \cite{wasp3a} which suggested an upper limit of 3.5 Gyr. We cannot put a lower limit on the age of the star using Figure 3 as the lower range of the errors lie on or beyond the Zero Age Main Sequence. The evolutionary traks of \cite{siess2000} suggest a pre-main sequence lifetime of 0.13 Gyr for a 1.2 M$_{\odot}$ star. WASP-3 is a main sequence star, so we can impose this value as an extreme lower age limit. In the discovery paper a lower age estimate of 0.7 Gyr is presented.
We can also determine an age estimate from gyrochronology. If we assume that the spin axis of WASP-3 is nearly perpendicular to
the line of sight, then $v\sin i = 13.9$ km s$^{-1}$ and $R_s = 1.21 R_
\odot$ yields
a spin period of  $4.5\sin i$ days. The period will be shorter than  
4.5 days if the inclination
is substantially less than 90 degrees.

WASP-3 has 2MASS colour J-K = 0.242.
The fastest rotators at the same colour in the 590-Myr-old Coma  
Berenices open cluster
have periods of order 6 days \citep{cameron2009b}.
Hyades stars of similar colour also have periods in the range 5-6 days  
at age 625 Myr
\citep{radick1987}, as is also found to be the case in  
the
Praesepe cluster (age 580 Myr) by Delorme et al (2010, MNRAS,  
submitted).

If we assume that WASP-3 is spinning down because of angular momentum  
loss
in a hot, magnetically-channeled stellar wind, its spin period should  
increase
with time as the square root of its age. From their calibration of the  
period-colour
relation in Hyades and Praesepe, Delorme et al (2010) find

$$
t = 625\left(\frac{P_{\rm rot}}{11.401+12.652(J-K-0.631)}\right)^2\  
\mbox{\rm Myr},
$$

yielding a gyrochronological age of 300 Myr for WASP-3, with an  
uncertainty of order
10 percent. We caution, however, that this relation is only applicable
once the spin rate has converged to the asymptotic period-colour  
relation.
While convergence is probably complete for most stars of this mass by  
the age
of 300 Myr, there remains a small possibility that the star could have  
been born
as a relatively slow rotator, leading to over-estimation of the  
gyrochronological
age. We can, however, state with confidence that the gyrochronological  
age of WASP-3
is substantially less than that of the Hyades.

\section{Conclusions}
The Rossiter-McLaughlin effect present in the WASP-3 system was analysed
from observations made using the SOPHIE spectrograph on the 1.93m telescope at the
Observatoire de Haute-Provence \citep{simpson2010}. We analysed the observations using a new method
developed by \cite{cameron2009} which involves decomposing the CCF into its various components
and directly tracking the trajectory of the missing starlight across the line profile.
This method was incorporated into an MCMC analysis of all photometric and spectroscopic
data available for the WASP-3 system and was found to produce good results for the system
parameters, helping further constrain the values of the spin-orbit
misalignment angle, projected stellar rotation rate and the impact parameter.
The value we obtained for the projected spin-orbit misalignment angle $\lambda = 9_{-5}^{+6}$$^\circ$ is
close to zero and agrees with the previous results found by \cite{simpson2010} and \cite{tripathi2010}. Our value of $v\sin i=13.9\pm{0.03}$ km s$^{-1}$
is in agreement with the value obtained from the spectroscopic broadening but determined to a much higher level of precision.
We conclude that this new method of analysing the Rossiter-McLaughlin effect
successfully retrieves a more accurate and precise value for the
projected stellar spin rate compared with previous measurements, and 
in doing so it is not vulnerable to the systematic error present in previous
methods that require fitting of Gaussians to non-Gaussian CCFs in order to
calculate the velocity shifts. It also finds a value for the spin-orbit misalignment angle that agrees with all previous measurements and is of a similar precision.
The fact that we clearly detect the signature of the missing light after subtracting the model out-of-transit profile
shows that the method works well for a stellar
rotation rate that is not much greater than the intrinsic line width. In fact, \cite{cameron2009} showed that this method can be successful when $v\sin i$ is as low as half the value of the intrinsic stellar line width.

The value we obtained for the spin-orbit alignment angle is small enough to support
the theory of migration through tidal interaction with a protoplanetary disc. However, recent
discoveries of highly misaligned systems \citep{wasp17, hatp7} suggest that migration cannot
be explained in all cases purely by tidal interactions in a protoplanetary disc. \cite{hebrard2010} showed that
planets with measured spin-orbit angles could be sorted into three distinct populations: firstly, the majority
of hot jupiters that are aligned; secondly, the few that are strongly misaligned; and thirdly, the massive planets
that are mostly moderately but significantly misaligned. They propose that this could be the signature of three distinct evolution scenarios. 
Measuring the distribution of alignment angles in Hot Jupiter systems and will help 
inform theories of planetary migration.

Many transit candidates are rejected from RV follow-up observations if they are found to be rotating too fast 
for high-precision determination of the RV shift. This is especially true for stars of spectral type earlier than \~F5 which tend to be line-poor rapid rotators.
It has already been shown \citep{wasp33} that using this new method we can successfully confirm the existence of a planet-sized object by measuring the size of the 'bump' on the stellar line profile,
opening up a way for establishing the existence of close-orbiting planets around rapidly-rotating stars previously inaccessible to planet hunters using the RV method. With this new information we will also be able to investigate any differences in planet formation between solar-type and early-type stars.

\begin{acknowledgements}
The authors of this paper would like to thank the team at the Observatoire de Haute Provence
for their support during the observing runs. We also thank the anonymous
referee for their helpful and constructive comments. This work is supported by the UK Science and Technology Facilities
Council and made use of the ADS database and VizieR catalogue.
G.R.M. Miller would like to acknowledge the support and advice given by
colleagues during the writing of his first paper.
\end{acknowledgements}

\bibliographystyle{aa}
\bibliography{refs}

\end{document}